\documentclass[conference]{IEEEtran}
\usepackage{cite}
\usepackage[pdftex]{graphicx}
\usepackage{amsmath}
\usepackage{algorithmic}
\usepackage{array}
\usepackage{caption} 
\usepackage{amsmath}
\newtheorem{theorem}{Theorem}

\begin{document}
\IEEEoverridecommandlockouts

\title{On asymptotically optimal   tests for random number generators}
\author{\IEEEauthorblockN{Boris Ryabko}
\IEEEauthorblockA{Institute of Computational Technologies of SB RAS\\ 
Novosibirsk state university
\\Novosibirsk, Russian Federation \\Email: boris@ryabko.net\\}
}

\maketitle

\begin{abstract}
The problem of constructing effective statistical tests for random number generators (RNG) is considered. Currently, statistical tests for RNGs are a mandatory part of cryptographic information protection systems, but their effectiveness is mainly estimated based on experiments with various RNGs.
 We 
 find an asymptotic estimate for the p-value of an optimal test in the case where the alternative hypothesis is a known stationary ergodic source, and then describe  
a family of tests each of  which has the same asymptotic estimate of the p-value 
for any (unknown) stationary ergodic source.

\end{abstract}

\textbf{keywords:}   randomness testing, statistical test for cryptographic applications,
random number generators, Shannon entropy.

\section{Introduction}

Random number generators (RNG) and pseudo-random number generators (PRNG) are widely used in many information technologies, including cryptographic information protection systems.    The goal of RNG and PRNG is to generate sequences of binary digits, which are distributed as a result of throwing an '' honest '' coin or, more precisely, obey the Bernoulli distribution with parameters $ (1/2, 1/2) $. As a rule, for practically used RNG and PRNG this property is verified experimentally with the help of statistical tests intended for this purpose, see for a review \cite{l2017history,l2007testu01}. 
 Nowadays, there are more than a hundred applicable statistical tests
 and some  of them
is a mandatory part of cryptographic information protection systems 
  \cite{NIST-test}. Besides,   there are   dozens RNGs based on different physical processes \cite{grev}, and an even greater number of PRNGs based on different mathematical algorithms;  see for review \cite{l2017history}.  In such a situation, the natural question is how to compare different tests? Currently, the main method of such a comparison is numerical experiments in which different tests are practically applied to different RNGs, see for review \cite{l2017history,l2007testu01,NIST-test}.

Here we consider the problem of finding optimal tests in the case when the RNG is modeled by stationary ergodic sources. We found the following asymptotic solution to this problem: we first described the asymptotic behaviour of the p-value of the optimal test for the case where the probability distribution of the RNG is a priori known, and then described a family of statistical tests that have the same asymptotic estimates of the p-value for any distribution (which is not known in advance). More precisely, we showed that in both cases, with probability 1, $ \lim_ {n \to \infty} - \frac {1} {n} \log \pi_\tau(x_1 x_2 ... x_n) = 1 - h (\nu) $, where $ x_1 x_2 ... x_n$ is the sample, $\tau$ is the test,  $ \pi_\tau(x_1 x_2 ... x_n) $ is the p-value, and $ h (\nu) $ is the Shannon entropy of the (unknown) RNG distribution  $ \nu $.   
It turns out that asymptotically optimal tests with the required properties 
 are known \cite{R1,R2},  and are deeply connected with  so-called universal codes. 
 Note
that 
nowadays there are many universal codes which 
are based on different ideas and approaches, among which we note the PPM universal code \cite{cleary1984data},  the arithmetic code \cite{rissanen1979arithmetic}), the Lempel-Ziv (LZ)  codes
\cite{ziv1977universal}, the Burrows-Wheeler transform \cite{burrows1994block} which is used along with the book-stack (or MTF) code \cite{ryabko1980data,bentley1986locally,ryabko1987technical},  
the  class of grammar-based codes \cite{kieffer2000grammar,yang2000efficient}
and some others \cite{drmota2010tunstall,BRyabko:84,willems1995context}. 
All these codes are universal. This means that, asymptotically, the length of the compressed file goes to the  smallest possible value,  i.e. the 
Shannon entropy  ($h(\nu)$) per letter. 

 The main idea of  randomness tests  based on universal codes is rather natural: try to ``compress'' a test sequence 
 by a universal code: if the sequence is significantly compressed, then it is not random, see
 \cite{R1,R2} and a short description below.

 The rest of the paper is organised as follows. The next section is contains the necessary  definitions  and some 
 basic facts used in the following. Sections~\ref{s3}, \ref{s4} are  devoted to investigation 
 of the  Neyman-Pearson test and tests  based on universal codes, correspondingly.
 The proofs are given in the appendix.

\section{ Definitions}
\subsubsection{The main notations}
We consider RNG which generates a sequence of letters $x = x_1 x_2 ... x_n, n \ge 1,$  from the alphabet  $ \{0,1\}^n$. There are two following  statistical hypothesis  $H_0 = $ 
$\{ x \, $ \emph{ 
obeys uniform distribution  ($ \mu_U$) on  
  $\{0,1\}^n$ \}}  and the alternative hypothesis $H_1 = \bar{H_0} 	$, 
that is, $H_1$ is negation of $H_0$. 
Let $T$ be a test. 
Then,  by definition, a significance level $\alpha$ equals probability 
of the Type I error.  (Recall, that Type I error occurs   if
$H_0$ is true and  $H_0$ is  rejected. Type II error occurs   if
$H_1$ is true, but $H_0$ is accepted.) 
Denote a critical region of the test $T$ for the  significance level $\alpha$ by $C_T(\alpha)$  and let 
$\bar{C}_T(\alpha)$ $ =  \{0,1\}^n \setminus C_T(\alpha) .$ 
  (Recall, that   
for a certain 
$x = x_1 x_2 ... x_n\,$ the hypothesis 
$H_0$ is rejected if and only if $x \in C_T(\alpha)$.)

 Let us assume that $H_1$ is true and  the investigated sequence 
 $x = x_1 x_2 ... x_n$ is generated by
(unknown)   source $\nu$.
By definition, the test $T$ is consistent (for  $\nu$), if for any 
significance level $\alpha \in (0,1)$ the probability of Type II error goes to 0, that is 
 \begin{equation}\label{cons}
 \lim_{n \to \infty} \nu( \bar{C}_T(\alpha) ) = 0 \, .
 \end{equation}

Let us give a definition of a
 so-called p-value, which  plays an important rule in the randomness testing. 
Let there be a   statistic $\tau$  (that is, a function on $\{0,1\}^n$) and $x$ be a word 
from $\{0,1\}^n$. 
A  p-value  ($\pi_\tau (x)$) of  $\tau$ and  $x$  is defined by the equation
 \begin{equation}\label{pi-tau}
\pi_\tau (x) = \mu_U \{ y: \tau(y) \ge \tau(x) \} = | \{ y: \tau(y) \ge \tau(x) \} | / 2^n \, .
 \end{equation}
(Here and below $|X|$ is a number of elements $X$, if $X$ is a set, and the length of $X$, if $X$ is a word.)  
 
 Informally, 
$ \pi_{\tau} (x)$ is the probability to meet a random point $y$ which is 
worse than the
observed when considering the null hypothesis.
\subsubsection{The consistent tests for stationary ergodic sources and universal codes}
Firs let us  give a short informal description of the  universal codes. 
For any integer $m$ a lossless  code $\phi$ is defined  as such a map  from the set
of $m$-letter words to the set of all binary words that for any sequence of 
encoded 
$m$-letter words 
$\phi(v_1) \phi( v_2) ...$  the initial sequence $v_1 v_2 ...$ can be found without 
mistakes; the formall definition can be found, for example, in \cite{co}.

We will consider so-called universal codes which have the   following property:
for any stationary ergodic $\nu$  defined on the set of all infinite binary words
$x   = x_1 x_2 ... $, with probability one
 \begin{equation}\label{un-lim}
\lim_{n \to \infty } \frac{1}{n} | \phi( x_1 x_2 ... x_n) | / n  = h(\nu) \, ,
\end{equation}
where $ h(\nu) $ is the Shannon entropy of $\nu$ (see for definition \cite{co}). Such codes exist,  see, for example, \cite{R1, R2}.
Note, that a goal of  codes is to " compress " sequences, i.e. make a  length of the codeword 
$\phi( x_1 x_2 ... x_n) $ as small as possible. The   property (\ref{un-lim}) shows
that the universal codes are asymptotically optimal, because the Shannon entropy is 
a low bound of the length of the compressed sequence (per letter), see \cite{co}.

Let us back to considered problem of hypothesis testing. 
Suppose, it is known that a sample sequence $x= x_1 x_2 ...$  was  generated by stationary ergodic source and, as before,  we consider the same $H_0$ against the same $H_1$. Let $\phi$ be a universal code. 
The following test is a particular case of a goodnes-of-fit test suggested in \cite{R1,R2}:

{\it
If 
 $n -  |\phi (x_1 ... x_n)| \ge - \log_2 \alpha $ 
  then $H_0$ is rejected, 
otherwise accepted.
Here, as before, $\alpha$ is the significance level,  $|\phi (x_1 ... x_n)|$  is the length 
of encoded (''compressed") sequence. }

We denote  this test by $T_\phi$ and its statistic  by $\tau_\phi$, i.e. 
 \begin{equation}\label{tau-fi}
\tau_\phi (x_1 ... x_n) = n - |\phi(x_1 ... x_n)| \, .  
\end{equation}

It turns out that this test is consistent for any stationary ergodic source. 
More precisely, the following theorem  is proven in \cite{R1,R2 }:

{\it  For each stationary ergodic $\nu$, $\alpha \in (0,1) $ and a universal code $\phi$, the Type I error of the described test is not larger than  $\alpha$ and  the Type II error goes to 0, when $n \to \infty$. }

\section{ Asymptotic behaviour of a p-value of the  Neyman-Pearson test.}\label{s3}

Suppose, that $H_1$ is true and sequences $x \in \{0,1\}^n$ are obey a certain distribution $\nu$.
It is well-known in mathematical statistics that  the optimal test  ($NP$-test or likelihood-ratio test)  is described by 
Neyman–-Pearson lemma and the critical region of this test  is defined 
as follows:
$$
C_{NP}(\alpha) = \{x :  \, \, \mu_U (x) / \nu (x) \le \lambda_\alpha  \} \, ,
$$
where $\alpha \in (0,1) $ is the 
significance level and the constant $\lambda_\alpha$ is chosen in such a way that 
$\mu_U ( C_{NP}(\alpha) ) = \alpha$, see \cite{ks }.
(We did not take into account that the set $ \{0,1\}^n$ is finite. 
Strictly speaking, in such a case a randomized test should be used, but in what follows 
we will consider asymptotic behaviour of the tests for large $n$ and this effect will be negligible).
The p-value for the $NP$-test can be derived from the definition (\ref{pi-tau}), if we put 
$\tau(x) = \nu(x)$ and take into account that by definition, $ \mu_U (x)  = 2^{-n}$ for any $x$ $\in \{0,1\}^n$. So, 
 \begin{equation}\label{pi-np}
\pi_{NP} (x) = \mu_U \{ y: \nu(y) \ge \nu(x) \} = | \{ y: \nu(y) \ge \nu(x) \} | / 2^n \, .
\end{equation}
The following theorem  describes an asymptotic behaviour of p-values for stationary ergodic sources for $NP$ test. 
\begin{theorem}\label{t1} 
 If $\nu$ is a stationary ergodic measure, then, with probability 1, 
\begin{equation}\label{t11}
 \lim_{n \to \infty} -  \frac{1}{n} \log \pi_{NP} (x) = 1 - h(\nu) \, ,
\end{equation}
where $h(\nu)$ is the Shannon entropy of $\nu$, see for definition \cite{co}.
\end{theorem}
The $NP$-test is optimal in the sense that its probability of a Type II error is minimal, but when testing RNG the alternative distribution is unknown, and, hence,  some other tests should be used.    It turns out that the above described test $T_\phi $ has the same  asymptotic
behaviour as $NP$-test.
\section{ Asymptotically optimal tests for randomness.}\label{s4}

The following theorem  describes an asymptotic behaviour of p-values for stationary ergodic sources for  tests which are based on universal codes.   
\begin{theorem}\label{t1} 
Let $\phi$ be a universal code and the test $T_\phi$ with statistic $\tau_\phi$ (\ref{tau-fi})  is applied.
Then
for   any  stationary ergodic measure  
$\nu$, with probability 1, 
\begin{equation}\label{t2}
 \lim_{n \to \infty} -  \frac{1}{n} \log \pi_{\tau_\phi} (x) = 1 - h(\nu) \, ,
\end{equation}
where $\pi_{\tau_\phi}$ is the p-value.
\end{theorem}
Note  that this theorem gives some idea of the relation between the Shannon entropy of the (unknown) process $ \nu $ and the required sample size.
Indeed,  suppose that a $ NP $ test is used and the desired significance level is $ \alpha $. Then, we can see  that
(asymptotically) $\alpha$ should be less than $\pi_{NP}(x)$ and from (\ref{t11}) 
we obtain 
$ n > - \log \alpha / (1-h(\nu) ) \, $ (for the most powerful  test).
It is known that the Shannon entropy is 1 if and only if $\nu$ is the uniform measure $\mu_u$.  Therefore, in a certain sense, the difference $ 1-h (\nu) $ estimates the distance between the distributions, and the last inequality shows that the required 
 sample size 
 goes to infinity if $ \nu $ approaches  the uniform distribution. 

The next simple example illustrates  the theorems. 
 Let there be a test $\kappa$ and a generator (a measure $\nu$)  that generates sequences of independent binary digits with, say,   $\nu(0) = 0.501, \nu(1) = 0.499 $.
Suppose that 
 $\lim_{n \to \infty} -  \frac{1}{n} \log \pi_{\kappa} (x) = c \, $,
 where $c$ is a positive constant.  Let us consider the following  ``decimation test''  $\kappa^{1/2}$:  an input sequence $x_1 x_2 .... x_n$ is transformed into $x_1 x_3 x_5 ... x_{2 \lfloor n/2 \rfloor -1} $ and then $\kappa$ is applied to this transformed sequence. Obviously, for this test 
 $\lim_{n \to \infty} -  \frac{1}{n/2} \log \pi_{\kappa^{1/2}} (x) = c  \, $, and, 
 hence, $\lim_{n \to \infty} -  \frac{1}{n} \log \pi_{\kappa^{1/2}} (x) = c /2 \, $. 
 Thus, the value $ - \frac {1} {n} \log \pi_{\kappa} (x_1 ... x_n) $ seems to be a reasonable estimate of the power of the test   for  large $n$.
 

\section{Appendix }
{\it Proof of Theorem 1.}   
 The well-known  Shannon-McMillan-Breiman (SMB) theorem   states that for the stationary ergodic source
 $\nu$ and any $\epsilon >0, \delta > 0$ there exists such $n'(\epsilon, \delta)$ that 
$$
 \nu \{  x\in \{0,1\}^n  : \,  \, h(\nu) - \epsilon < - \frac{1}{n } \log \nu(x)   <  $$ 
  \begin{equation}\label{smb}
 h(\nu) + \epsilon \, \, \} >1 - \delta \,  \quad  for \, \, \,  n > n'(\epsilon,\delta) \, , 
 \end{equation} 
 see \cite{co}.  From this we obtain 
$$ \nu \{   x \in \{0,1\}^n  : \, \,   2^{- n (h(\nu) -  \epsilon )} >\, \nu(x) > 
$$  \begin{equation}\label{smb2}
 2^{- n (h(\nu) +  \epsilon )}  \}
 >1 - \delta \, 
 \end{equation}
  for $n > n'(\epsilon, \delta)$. 
  It will be convenient to define
  $$
  \Phi_{\epsilon, n} = \{ x\in \{0,1\}^n  :  \, h(\nu) - \epsilon < 
   $$ \begin{equation}\label{fi}   - \frac{1}{n } \log \nu(x)   <
 h(\nu) + \epsilon \, \, \}  
 \end{equation} 
  From this definition and (\ref{smb2} )  we obtain 
 \begin{equation}\label{f2}
(1-\delta) \, 2^{n (h(\nu) -\epsilon) } \le | \Phi_{\epsilon, n} | \le 2^{n (h(\nu) +\epsilon) }\, .
 \end{equation} 
For any $x \in \Phi_{\epsilon,  n}$ define 
 \begin{equation}\label{lam}
\Lambda_x  = \{ y: \nu (y) \ge \nu(x) \, \, \}  \bigcap  \Phi_{\epsilon,  n} \, .
 \end{equation}  Note that, by definition, $|\Lambda_x| \le | \Phi_{\epsilon, n}|$
 and from (\ref{f2}) we obtain
 \begin{equation}\label{lam2} | \Lambda_x | \le 2^{n (h(\nu) +\epsilon) }\, . \end{equation} 
For any $\rho \in (0,1)$ we define $\Psi_\rho \subset \Phi_{\epsilon, n}$ such that
 \begin{equation}\label{psi}
 \nu( \Psi_\rho) = \rho \,\, 
\&  \,  \, \forall 	u \in \Psi_\rho ,\,    \forall 	v \in ( \Phi_{\epsilon,  n} \setminus 
\Psi_\rho )\, : 
 \nu(u) \ge \nu(v) \, .
 \end{equation} 
 (That is, $\Psi_\rho $ contains the most probable words whose total probability equals $\rho$.  If there are several such sets we can take any of them. )
 Let us consider any $x \in ( \Phi_{\epsilon,  n} \setminus 
\Psi_\rho )  $.  Taking into account 
the definition (\ref{psi}) and (\ref{f2})
we can see that for this $x$
  \begin{equation}\label{lval-}
|\Lambda_x | \ge \rho | \Phi_{\epsilon, n}| \ge \rho (1-\delta) 2^{n (h(\nu) -\epsilon) } \, .
 \end{equation}  
So, from this inequality and (\ref{lam2}) we obtain 
\begin{equation}\label{lam-size}
\rho (1-\delta) 2^{n (h(\nu) -\epsilon) } \le |\Lambda_x | \le \,2^{n (h(\nu) +\epsilon) } \, . \end{equation}   
From equation  (\ref{smb2}), (\ref{fi}) and (\ref{psi}) we can see that 
$\nu ( \Phi_{\epsilon, n} \setminus \Psi_\rho ) \ge (1-\delta) (1-\rho) $.
Taking into account (\ref{lam-size}) and this inequality, we can see that 
$$ \nu \{x:  \, 
 h(\nu) - \epsilon + \log ( \rho (1-\delta))/ n 
 $$ 
  \begin{equation}\label{fin}
\le  \log
|\Lambda_x | /n  \le h(\nu) +  \epsilon \} \ge  (1-\delta) (1-\rho) . 
 \end{equation}  
 From the definition (\ref{pi-np}) of $\pi_{NP}(x)$ and the definition
 (\ref{lam}) of $\Lambda_x $, we 
can see that $ \pi_{NP}(x) = |\Lambda_x |/ 2^n$.
Taking into account this equation and (\ref{fin})
 we obtain  the following:
 $$  \nu \{x: \,1 - ( h(\nu) - \epsilon +\log  ( \rho (1-\delta)) / n ) \ge $$
 \begin{equation}\label{fin2} 
  - \log
\pi_{NP}(x) /n  \ge 1 - ( h(\nu) +  \epsilon )  \} \ge  (1-\delta) (1-\rho) . 
 \end{equation}  
Clearly, there exists such $n^*(\rho)$ that for $n >n^*(\rho)$  $\, \,\,\, \, - \log  ( \rho (1-\delta)) / n
< \epsilon$. Taking into account (\ref{smb}) we can see that  
$$  \nu \{x: \,1 - ( h(\nu) - 2 \epsilon)  \ge $$
\begin{equation}\label{fin2} 
  - \log
\pi_{NP}(x) /n  \ge 1 - ( h(\nu) +  \epsilon )  \} \ge  (1-\delta) (1-\rho) 
\end{equation}  
for $n > \max (n'(\epsilon, \delta), n^*(\rho))$. 
This inequality is valid for any $\rho \in (0,1)$ and, in particular, for $\rho = \delta$. 
So, from (\ref{fin2}) we obtain 
$$  \nu \{x: \,1 - ( h(\nu) - 2 \epsilon ) \ge $$
$$
  - \log
\pi_{NP}(x) /n  \ge 1 - ( h(\nu) +  \epsilon )  \} \ge  (1- 2 \delta) . 
$$
for $n > \max (n'(\epsilon, \delta), n^*(\delta))$. 

Having taken into account that this inequality is valid for all positive $\epsilon$ and $ 
 \delta$,  we obtain the  statement of the theorem.

{\it Proof of Theorem 2}  is similar to the previous one.  First, for any
$\epsilon >0, \delta >0$ 
 we define 
 \begin{equation}\label{abc}
 \hat{\Phi}_{\epsilon, n} = \{x: h(\nu) - \epsilon < | \phi(x_1 ... x_n) | /n <  h(\nu) + \epsilon \, \}\, .
 \end{equation}  
Note that
from (\ref{un-lim} ) we can see that there exists  such $n''(\epsilon,\delta)$ that,
for $n > n''(\epsilon,\delta)$, 
\begin{equation}\label{fi-hat}
\nu ( \hat{\Phi}_{\epsilon, n})  > 1 - \delta \, .
 \end{equation}  
 We will use the set $\Phi_{\epsilon,  n}$ (see (\ref{fi}) ). 
 Having taken into account 
the  SMB theorem (\ref{smb}) and   (\ref{fi-hat}), 
we can see that 
\begin{equation}\label{fi-hat2}
\nu ( \hat{\Phi}_{\epsilon,  n} \cap    \Phi_{\epsilon,  n} )    > 1 - 2 \delta \, ,
 \end{equation}  
if $n > \max (n'(\epsilon,\delta), n''(\epsilon,\delta))$. 

From this moment, the proof begins to repeat the proof of the first theorem,
 if we use the set $(\hat{\Phi}_{\epsilon, n} \cap    \Phi_{\epsilon,  n} ) $ instead of  $ \Phi_{\epsilon,  n}$.  
 Namely, define
 \begin{equation}\label{La-new}
\hat{\Lambda}_x  = \{ y: |\phi (y) | \le |\phi(x)| \, \, \}  \cap  (\hat{\Phi}_{\epsilon,  n} \cap    \Phi_{\epsilon, n} ) \, 
 \end{equation}  
 and $ \hat{\Psi}_\rho $ is such a subset of 
$(\hat{\Phi}_{\epsilon,  n} \cap    \Phi_{\epsilon,  n} )$  that
$$
 \nu( \hat{\Psi}_\rho) = \rho \,\, 
\&  \,  \, \forall 	u \in \Psi_\rho \, ,   \forall 	v \in ( (\hat{\Phi}_{\epsilon,  n} \cap    \Phi_{\epsilon,  n} ) \setminus 
\Psi_\rho ):\,  
$$
\begin{equation}\label{psi-new}
 |\phi(u) |\le |\phi(v) | \, .
 \end{equation} 
 Let us consider any $x \in (  (\hat{\Phi}_{\epsilon,  n} \cap    \Phi_{\epsilon, n} )\setminus 
\hat{\Psi}_\rho )  $.  
Taking into account 
the definition (\ref{La-new})   and   (\ref{fi-hat2}),
 we obtain 
\begin{equation}\label{lam-size-new}
\rho (1- 2 \delta) 2^{n (h(\nu) -\epsilon) } \le | \hat{\Lambda}_x | \le \,2^{n (h(\nu) +\epsilon) } \,. 
\end{equation}  
From equations  (\ref{fi-hat2}) and (\ref{psi-new}) we can see that 
$\nu ( (\hat{\Phi}_{\epsilon,  n} \cap    \Phi_{\epsilon, n} ) \setminus \hat{\Psi}_\rho ) \ge (1- 2 \delta) (1-\rho) $.
Taking into account (\ref{lam-size-new}) and this inequality, we can see that 
$$ \nu \{x:  \, 
 h(\nu) - \epsilon + \log ( \rho (1-2 \delta))/ n 
 $$ 
  \begin{equation}\label{fin-hat}
\le  \log
|\hat{\Lambda}_x | /n  \le h(\nu) +  \epsilon \} \ge  (1-2 \delta) (1-\rho) . 
 \end{equation}  
 From the definition   of p-value (\ref{pi-tau})  and the definition
 (\ref{La-new}), we 
can see that $ \pi_{\tau_\phi}(x) = |\hat{\Lambda}_x |/ 2^n$.
Taking into account this equation and (\ref{fin-hat})
 we obtain  the following:
 $$  \nu \{x: \,1 - ( h(\nu) - \epsilon +\log  ( \rho (1-\delta)) / n ) \ge $$
 \begin{equation}\label{fin2-hat} 
  - \log
\pi_{\tau_\phi}(x) /n  \ge 1 - ( h(\nu) +  \epsilon )  \} \ge  (1-2 \delta) (1-\rho) . 
 \end{equation}  
Clearly, there exists such $n^{**}(\rho)$ that for $n >n^{**}(\rho)$  $\, \,\,\, \, - \log  ( \rho (1- 2 \delta)) / n
< \epsilon$. 
Taking it account  we can see from (\ref{fin2-hat}) that  
$$  \nu \{x: \,1 - ( h(\nu) - 2 \epsilon)  \ge $$
\begin{equation}\label{fin2-hat2} 
  - \log
\pi_{\tau_\phi}(x) /n  \ge 1 - ( h(\nu) +  \epsilon )  \} \ge  (1-2 \delta) (1-\delta) 
\end{equation}  
for $n > \max (n'(\epsilon, \delta), n''(\epsilon, \delta),n^{**}(\rho))$. 
So, from (\ref{fin2-hat2}) we obtain 
$$  \nu \{x: \,1 - ( h(\nu) - 2 \epsilon ) \ge $$
$$
  - \log
\pi_{\tau_\phi}(x) /n  \, \ge  \, \, 1 - ( h(\nu) +  \epsilon )  \} \ge  (1- 3 \delta) . 
$$
for $n > \max (n'(\epsilon, \delta), n''(\epsilon, \delta),n^{**}(\delta))$. 

Having taken into account that this inequality is valid for all positive $\epsilon$ and $ 
 \delta$,  we obtain the  statement of the theorem.
\section*{Acknowledgment}
Research  was supported  by  Russian Foundation for Basic Research
(grant no. 18-29-03005).


\begin{thebibliography}{10}

\bibitem{l2017history}
L'Ecuyer, P. History of uniform random number generation. In Proceedings of the WSC
  2017-Winter Simulation Conference, Las Vegas, NV, USA, 3--6 December 2017.
  
  
\bibitem{l2007testu01}
P.~L'Ecuyer and R.~Simard, ``TestU01: AC library for empirical testing of random number generators,'' {\em ACM Transactions on Mathematical Software}, vol. 33, no. 4, p.22, 2007.



\bibitem{NIST-test}
A.~Rukhin, J.~Soto, J.~Nechvatal, M.~Smid, E.~Barker, S.~Leigh, M.~Levenson,
  M.~Vangel, D.~Banks, A.~Heckert, J.~Dray, and S.~Vo, {\em A Statistical Test
  Suite for Random and Pseudorandom Number Generators for Cryptographic
  Applications}.
\newblock National Institute of Standards and Technology, 2010.



\bibitem{grev}
Herrero-Collantes, M., Garcia-Escartin, J.C. Quantum random number generators,
{\em  Rev. Mod. Phys.}, vol. 89, ~015004, 2017.


\bibitem{R1}
B.  Ryabko and  J. Astola,  `` Universal Codes as a Basis for Time Series Testing,"
\emph{  Statistical Methodology}, vol.3, pp. 375-397, 2006.

\bibitem{R2}
B. Ryabko, J.  Astola, M.  Malyutov,
 \emph{  Compression-Based Methods of Statistical Analysis and Prediction of Time Series,}
  Springer International Publishing Switzerland, 2016.





\bibitem{cleary1984data}
J.~Cleary and I.~Witten, ``Data compression using adaptive coding and partial
  string matching,'' \emph{IEEE Transactions on Communications}, vol.~32,
  no.~4, pp. 396--402, 1984.

\bibitem{rissanen1979arithmetic}
J.~Rissanen and G.~G. Langdon, ``Arithmetic coding,'' \emph{IBM Journal of
  research and development}, vol.~23, no.~2, pp. 149--162, 1979.

\bibitem{ziv1977universal}
J.~Ziv and A.~Lempel, ``A universal algorithm for sequential data
  compression,'' \emph{IEEE Transactions on information theory}, vol.~23,
  no.~3, pp. 337--343, 1977.

\bibitem{burrows1994block}
M.~Burrows and D.~J. Wheeler, ``A block-sorting lossless data compression
  algorithm,'' 1994.

\bibitem{ryabko1980data}
B.~Y. Ryabko, ``Data compression by means of a “book stack”,''
  \emph{Problems of Information Transmission}, vol.~16, no.~4, pp. 265--269, 1980.

\bibitem{bentley1986locally}
J. Bentley, D. Sleator,  R. Tarjan,  and V. Wei, 
`` A locally adaptive data compression scheme,"
\emph{Communications of the ACM},
  vol. 29,
  no. 4,
  pp. 320--330,
 1986.

\bibitem{ryabko1987technical}
B.~Ryabko, N.~R. Horspool, G.~V. Cormack, S.~Sekar, and S.~B. Ahuja,
  ``Technical correspondence,'' \emph{Communications of the ACM}, vol.~30,
  no.~9, pp. 792--797, 1987.

\bibitem{kieffer2000grammar}
J.~C. Kieffer and E.-H. Yang, ``Grammar-based codes: a new class of universal
  lossless source codes,'' \emph{IEEE Transactions on Information Theory},
  vol.~46, no.~3, pp. 737--754, 2000.

\bibitem{yang2000efficient}
E.-H. Yang and J.~C. Kieffer, ``Efficient universal lossless data compression
  algorithms based on a greedy sequential grammar transform. i. without context
  models,'' \emph{IEEE Transactions on Information Theory}, vol.~46, no.~3, pp.
  755--777, 2000.

\bibitem{drmota2010tunstall}
M. Drmota, Yu.  Reznik,   and W. Szpankowski, 
`` Tunstall code, Khodak variations, and random walks,"
 \emph{  IEEE Transactions on Information Theory},
  vol. 56,
  no. 6,
  pp. 2928--2937, 2010.
\bibitem{BRyabko:84}
B. Ryabko, ``Twice-universal coding,'' \emph{Problems of Information
  Transmission}, vol.~3, pp. 173--177, 1984.


\bibitem{willems1995context}
 F.M.J. Willems,  Yu.  Shtarkov,  and T.J.  Tjalkens, 
`` The context-tree weighting method: basic properties,"
   \emph{IEEE Transactions on Information Theory},
  vol. 41,
  no. 3,
  pp. 653--664, 1995.
  

\bibitem{co}
T.~M. Cover and J.~A. Thomas, \emph{Elements of information theory}.\hskip 1em
  plus 0.5em minus 0.4em\relax New York, NY, USA: Wiley-Interscience, 2006.

\bibitem{ks}
M. Kendall,  A. Stuart,   {\em The advanced theory of statistics; Vol.2:
  Inference and Relationship};
\newblock Hafner Publishing Company: New York, NY, USA, 
 1961.


\end{thebibliography}

\end{document}